\documentclass[reprint,pra,superscriptaddress,floatfix]{revtex4-1}

\usepackage{graphicx}
\usepackage{amsmath}
\usepackage{amssymb}
\usepackage{braket}
\usepackage[export]{adjustbox}

\usepackage{tabularx}
\usepackage{complexity}
\newclass{\SampP}{SampP}
\newclass{\SampBQP}{SampBQP}
\newclass{\IQP}{IQP}
\newclass{\BosonSampling}{BosonSampling}
\newclass{\SharpP}{\#P}

\begin{document}
\title{Quantum Sampling Problems, BosonSampling and Quantum Supremacy}

\newcommand{\affone}{Centre for Quantum Computation and Communications Technology, School of Mathematics and Physics, The University of Queensland, St Lucia, Queensland 4072, Australia}
\newcommand{\afftwo}{Centre for Quantum Computation and Communications Technology, Centre for Quantum Software and Information, Faculty of Engineering and Information Technology, University of Technology Sydney, NSW 2007, Australia}

\author{A.~P.~Lund}
\affiliation{\affone}
\author{Michael~J.~Bremner}
\affiliation{\afftwo}
\author{T.~C.~Ralph}
\affiliation{\affone}

\begin{abstract}
There is a large body of evidence for the potential of greater computational power using information carriers that are quantum mechanical over those governed by the laws of classical mechanics.  But the question of the exact nature of the power contributed by quantum mechanics remains only partially answered.  Furthermore, there exists doubt over the practicality of achieving a large enough quantum computation that definitively demonstrates quantum supremacy.  Recently the study of computational problems that produce samples from probability distributions has added to both our understanding of the power of quantum algorithms and lowered the requirements for demonstration of fast quantum algorithms.  The proposed quantum sampling problems do not require a quantum computer capable of universal operations and also permit physically realistic errors in their operation.  This is an encouraging step towards an experimental demonstration of quantum algorithmic supremacy.  In this paper, we will review sampling problems and the arguments that have been used to deduce when sampling problems are hard for classical computers to simulate.  Two classes of quantum sampling problems that demonstrate the supremacy of quantum algorithms are $\BosonSampling$ and $\IQP$ Sampling.  We will present the details of these classes and recent experimental progress towards demonstrating quantum supremacy in $\BosonSampling$.
\end{abstract}

\maketitle

\section{Introduction}

There is a growing sense of excitement that in the near future prototype quantum computers might be able to outperform any classical computer. That is, they might demonstrate supremacy over classical devices~\cite{Preskill12}. This excitement has in part been driven by theoretical research into the complexity of intermediate quantum computing models, which over the last 15 years has seen the physical requirements for a quantum speedup lowered while increasing the level of rigour in the argument for the difficulty of classically simulating such systems. 

These advances are rooted in the discovery by Terhal and DiVincenzo \cite{Terhal02} that sufficiently accurate classical simulations of even quite simple quantum computations could have significant implications for the interrelationships between computational complexity classes~\cite{Papadimitriou}. Since then the theoretical challenge has been to demonstrate such a result holds for levels of precision commensurate with what is expected from realisable quantum computers.  A first step in this direction established that classical computers cannot efficiently mimic the output of ideal quantum circuits to within a reasonable multiplicative (or relative) error in the frequency with which output events occur without similarly disrupting the expected relationships between classical complexity classes~\cite{BJS10, AA}. In a major breakthrough Aaronson and Arkhipov laid out an argument for establishing that efficient classical simulation of linear optical systems was not possible, even if that simulation was only required to be accurate to within a reasonable total variation distance. Their argument revealed a deep connection between the complexity of sampling from quantum computers and conjectures regarding the average-case complexity of a range of combinatorial problems.  The linear optical system they proposed was the class of problems called $\BosonSampling$ which is the production of samples from Fock basis measurements of linearly scattering individual Bosons.  Using the current state of the art of classical computation an implementation of $\BosonSampling$ using 50 photons would be sufficient to demonstrate quantum supremacy.

Since then many experimental teams have attempted to implement Aaronson and Arkhipov's $\BosonSampling$ problem~\cite{TIL13, SPR13, CRE13, BAR13, BRO13, SPA14} while theorists have extended their arguments to apply to a range of other quantum circuits, most notably commuting quantum gates on qubits, a class known as $\IQP$~\cite{BMS15}. These generalizations give hope that experimental demonstration of quantum supremacy on sufficiently high fidelity systems of just 50 qubits~\cite{Boixo16}. 

In this review we will present the theoretical background behind $\BosonSampling$ and its generalizations, while also reviewing recent experimental demonstrations of $\BosonSampling$. 
From a theoretical perspective we focus on the connections between the complexity of counting problems and the complexity of sampling from quantum circuits. This is of course not the only route to determining the complexity of quantum circuit sampling, and recent work by Aaronson and Chen explores several interesting alternative pathways \cite{Aaronson16}.

\section{Computational complexity and quantum supremacy}

\begin{table}
	\caption{\label{decision-table}Definitions of complexity classes used in this review. The ``Type'' column describes the outputs generated by algorithms within the class.  ``{\em D}'' denotes decision problems which output a single bit, whose values are often interpreted as 'accept' and 'reject'. ``{\em C}'' denotes counting problems which output a non-negative integer. ``{\em Z}'' denotes problems that generalise counting problems and allow negative integer outputs.  The definitions give the properties algorithms are required to have within each class. }
\begin{ruledtabular}
\begin{tabularx}{60mm}{ccX}
Class & Type & Definition \\ \hline
\P & {\em D} & deterministic with polynomial runtime on a classical computer \\
\EQP & {\em D} & deterministic with polynomial runtime on a quantum computer \\
\BPP & {\em D} & random with classical statistics and an error probability less than $1/3$  \\
\BQP & {\em D} & random with quantum statistics and an error probability less than $1/3$ \\
\NP & {\em D} & outputs can be verified using an algorithm from \P \\
\PP & {\em D} & random with classical statistics and an error probability less than $1/2$ \\
\SharpP & {\em C} & counts the number of 'accept' outputs for circuits from \P \\
\GapP & {\em Z} & difference between the number of 'accept' and 'reject' outputs for circuits from \P \\
\PSPACE & {\em D} & polynomial memory requirements on a classical computer
\end{tabularx}
\end{ruledtabular}
\end{table}
 
The challenge in rigorously arguing for quantum supremacy is compounded by the difficulty of bounding the ultimate power of classical computers. Many examples of significant quantum speedups over the best-known classical algorithms have been discovered, see \cite{Montanaro15} for a useful review. The most celebrated of these results is Shor's polynomial time quantum algorithm for factorisation~\cite{Shor95}.  This was a critically important discovery for the utility of quantum computing, but was not as satisfying in addressing the issue of quantum supremacy due to the unknown nature of the complexity of factoring.  The best known classical factoring algorithm, the general number field sieve, is exponential time (growing as $e^{c n^{1/3} \ln^{2/3}{n}}$ where $n$ is the number of bits of the input number). However, in order to prove quantum supremacy, or really any separation between classical and quantum computational models, it must proven for {\em all} possible algorithms and not just those that are known. 

The challenge of bounding the power of classical computation is starkly illustrated by the persistent difficulty of resolving the $\P$ versus $\NP$ question, where the extremely powerful non-deterministic Turing machine model cannot be definitively proven to be more powerful than standard computing devices. The study of this question has led to an abundance of nested relationships between classes of computational models, or complexity classes. Some commonly studied classes are shown in TABLE~\ref{decision-table}.  Many relationships between the classes can be proven, such as $\P \subseteq \NP$, $\PP \subseteq \PSPACE$ and $\NP \subseteq \PP$, however, strict containments are rare. Questions about the nature of quantum supremacy are then about what relationships one can draw between the complexity classes when introducing quantum mechanical resources.

A commonly used technique in complexity theory is to prove statements relative to an ``oracle''. This is basically an assumption of access to a machine that solves a particular problem instantly.  Using this concept one can define a nested structure of oracles called the ``polynomial hierarchy''\cite{Stockmeyer76} of complexity classes.  At the bottom of the hierarchy are the classes $\P$ and $\NP$ which are inside levels zero and one respectively.  Then there is the second level which contains the class $\NP^{\NP}$ which means problems solvable in $\NP$ with access to an oracle for problems in $\NP$.  If $\P \neq \NP$ then this second level is at least as powerful the first level and possibly more powerful due to the ability to access the oracle.  Then the third level contains $\NP^{\NP^{\NP}}$, and so on.  Higher levels are defined by continuing this nesting.  Each level of the hierarchy contains the levels below it.  Though not proven, it is widely believed that every level is strictly larger than the next.  This belief is primarily due to the relationships of this construction to similar hierarchies such as the arithmetic hierarchy for which higher levels are always strictly larger.  If it turns out that two levels are equal, then one can show that higher levels do not increase and this situation is called a polynomial hierarchy collapse.  A polynomial hierarchy collapse to the first level would mean that $\P=\NP$.  A collapse at a higher level is a similar statement but relative to an oracle.   It is the belief that there is no collapse of the polynomial hierarchy at any level that is used in demonstrating the supremacy of quantum sampling algorithms.  Effectively one is forced into a choice between believing that the polynomial hierarchy of classical complexity classes collapses or that quantum algorithms are more powerful than classical ones.

\section{Sampling problems}
\label{SecSampling}
\begin{figure}
	(a)
	\includegraphics[valign=t,width=80mm]{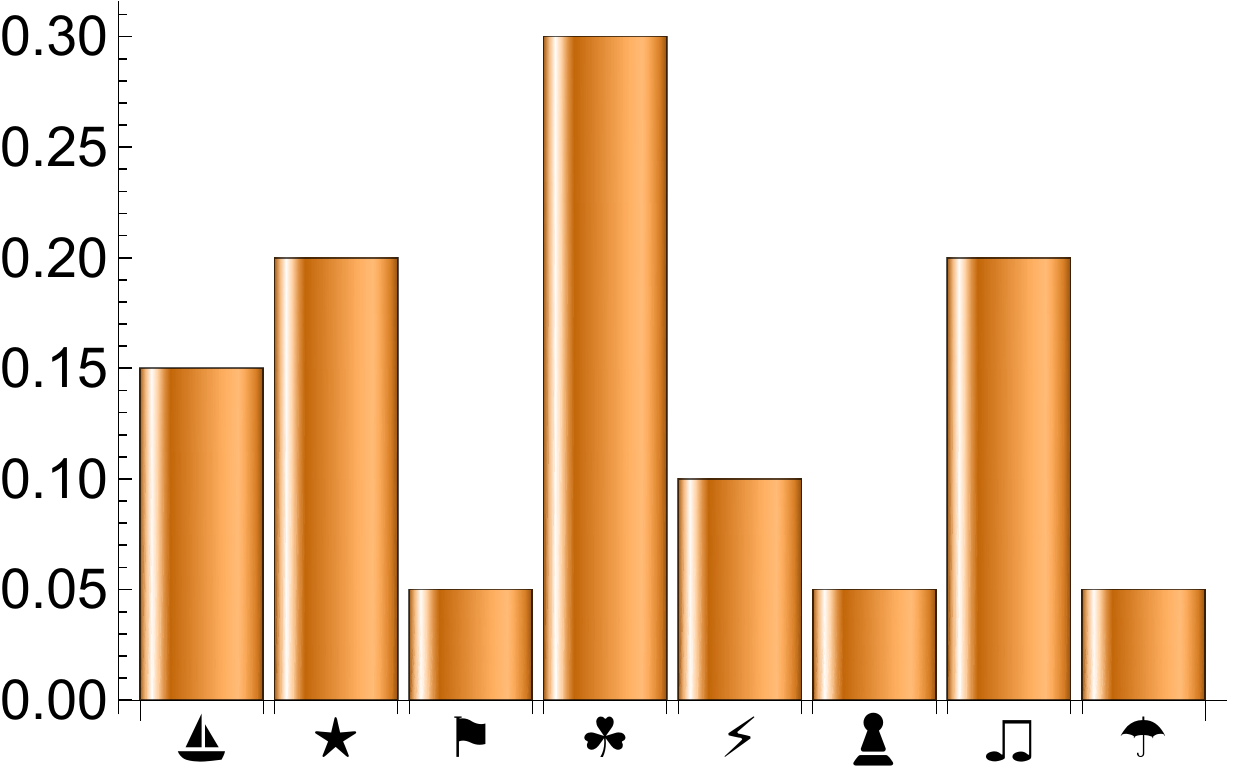} \\
	\vspace{4mm}
	(b)
	\includegraphics[valign=t,width=80mm]{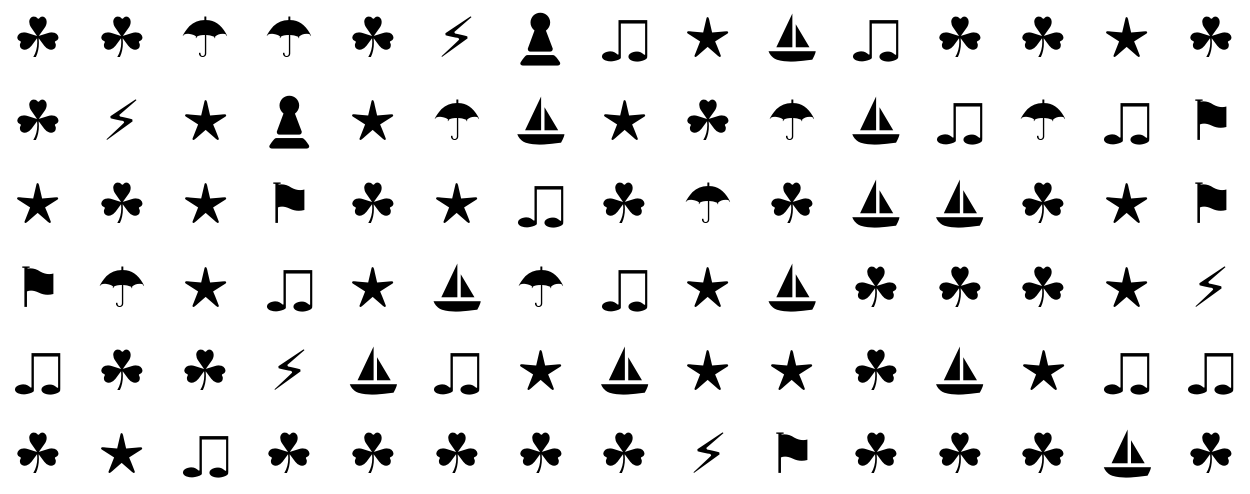}
	\caption{\label{SampDemo}(a) An example probability distribution over 8 symbols. (b) 100 random samples from the probability distribution.  The objective of a sampling problem is the compute samples like the sequences shown in (b) whose complexity may be different to the complexity of computing the underlying probability distribution (a).}
\end{figure}
Sampling problems are those problems which output random numbers according to a particular probability distribution (see FIG.~\ref{SampDemo}).  In the case of a classical algorithm, one can think of this class as being a machine which transforms uniform random bits into non-uniform random bits according to the required distribution.  When describing classes of sampling problems the current convention is to prefix ``Samp-'' to the class in which computation takes place.  So $\SampP$ is the class described above using an efficient classical algorithm and $\SampBQP$ would be those sampling problems which are efficiently computable using a quantum mechanical algorithm with bounded error.

All quantum computations on $n$ qubits can be expressed as the preparation of an $n$-qubit initial state $|0\rangle^{\otimes n}$, a unitary evolution corresponding to a uniformly generated quantum circuit $C$ followed by a measurement in the computational basis on this system. In this picture the computation outputs a length $n$ bitstring $x\in\{0,1\}^n$ with probability 
\begin{equation}
	\label{qu-prob}
	p_x=|\langle x|C|0\rangle^{\otimes n}|^2.
\end{equation}
In this way quantum computers produce probabilistic samples from a distribution determined by the circuit $C$. Within this model $\BQP$ is those decision problems solved with a bounded error rate by measuring a single output qubit. $\SampBQP$ is the class of problems that can be solved when we are allowed to measure all of the output qubits. 

It is known that quantum mechanics produces statistics which cannot be recreated classically as in the case of quantum entanglement and Bell inequalities.  However, these scenarios need other physical criterion to be imposed, such as sub-luminal signaling, to rule out classical statistics. Is there an equivalent ``improvement'' in sampling quantum probability distributions when using complexity classes as the deciding criterion?  That is, does $\SampBQP$ strictly contain $\SampP$? The answer appears to be yes and there is a (almost) provable separation between the classical and quantum complexity. 

Key to these arguments is understanding the complexity of computing the output probability of a quantum circuit from Eq.~(\ref{qu-prob}).  In the 1990s it was shown that there are families quantum circuits for which computing $p_x$ is $\SharpP$-hard in the \emph{worst-case} \cite{Fortnow98, Fenner98}.  The suffix ``-hard'' is used to indicate that the problem can, with a polynomial time overhead, be transformed into any problem within that class.  $\SharpP$-hard includes all problems in $\NP$.  Also, every problem inside the Polynomial Hierarchy can be solved inside the class of decision problems within $\SharpP$-hard, which is written $\P^\SharpP$~\cite{Toda}. Importantly this $\SharpP$-hardness does not necessarily emerge only from the most complicated quantum circuits, but rather can be established even for non-universal, or intermediate, families of quantum circuits such as $\IQP$ \cite{BJS10} and those used in $\BosonSampling$ \cite{AA}. This is commonly established by demonstrating that any quantum circuit can be simulated using the (non-physical) resource of postselection alongside the intermediate quantum computing model \cite{Terhal02,BJS10}.

In fact it is possible to show that computing $p_x$ for many, possibly intermediate, quantum circuit families is actually $\GapP$-complete, a property that helps to establish their complexity under approximations. $\GapP$ is a slight generalization of $\SharpP$ that contains all of the problems inside $\SharpP$ (see table \ref{decision-table}). Note that the suffix ``-complete'' indicates that the problem is both -hard and a member of the class itself. An estimate $\tilde{Q}$ of a quantity $Q$ is accurate to within a {\em multiplicative} error $\epsilon^\prime$ when $Q e^{-\epsilon^\prime} \leq \tilde{Q} \leq Q e^{\epsilon^\prime}$ or alternatively, as $\epsilon^\prime$ small is the usual case of interest, $Q (1 - \epsilon^\prime) \leq \tilde{Q} \leq Q (1 + \epsilon^\prime)$.  When a problem is $\GapP$-complete it can be shown that multiplicative approximations of the outputs from these problems are still $\GapP$-complete.  


It is important to recognize that quantum computers are not expected to be able to calculate multiplicative approximations to $\GapP$-hard problems, such as computing $p_x$, in polynomial time. This would imply that quantum computers could solve any problem in $\NP$ in polynomial time, which is firmly believed to not be possible. However, an important algorithm from Stockmeyer~\cite{Stockmeyer83} gives us the ability to compute good multiplicative approximations to $\SharpP$-complete problems by utilizing an $\NP$ oracle and by sampling from polynomial-sized classical circuits. The stark difference in complexity under approximations between $\SharpP$ and $\GapP$ can be used to establish a separation between the difficulty of sampling from classical and quantum circuits. If there were an efficient classical algorithm for sampling from families of quantum circuits with $\GapP$-hard output probabilities, then we could use Stockmeyer's algorithm to find a multiplicative approximation to these probabilities with complexity that is inside the third level of the Polynomial Hierarchy, however this causes a contradiction because $\P^\GapP$ contains the entire Polynomial Hierarchy (and it is assumed to not collapse). With such arguments it can be shown that it is not possible to even sample from the outputs to within a constant multiplicative error of many intermediate quantum computing models without a collapse in the Polynomial Hierarchy \cite{AA, BJS10, BJS10, Morimae13, Jozsa14, Bouland16}.


Such results suggest quantum supremacy can be established easily, however, quantum computers can only achieve \emph{additive} approximations to their own ideally defined circuits.  An estimate $\tilde{Q}$ of a quantity $Q$ is accurate within an additive error $\epsilon$ if $Q - \epsilon \leq \tilde{Q} \leq Q + \epsilon$.  Implementations of quantum circuits are approximate in an additive sense because of the form of naturally occurring errors, our limited ability to learn the dynamics of quantum systems, and finally because quantum circuits use only finite gate sets.  In order to demonstrate quantum supremacy we need a fair comparison between what a quantum computer can achieve and what can be achieved with classical algorithms. Following the above line of reasoning, we would need to demonstrate that if a classical computer could efficiently produce samples from a distribution which is close in an additive measure, like the total variation distance, from the target distribution then we would also see a collapse in the Polynomial Hierarchy. Being close in total variation distance means, with error budget $\beta$, samples from a probability distribution $q_x$ satisfying $\sum_x |p_x - q_x| \leq \beta$ are permitted.  An error of this kind will tend to generate {\em additive} errors in the outputs.  The key insight of Aaronson and Arkhipov was that for some special families of randomly chosen quantum circuits an overall additive error budget causes Stockmeyer's algorithm to give an additive estimate $\tilde{Q}$ that might \emph{also} be a good multiplicative approximation.

\section{$\BosonSampling$ problems}

Aaronson and Arkipov~\cite{AA} describe a simple model for producing output probabilities that are $\SharpP$-hard.  Their model uses bosons that interact only by linear scattering.  The bosons must be prepared in a Fock state and measured in the Fock basis.  

Linear bosonic interactions, or linear scattering networks, are defined by dynamics in the Heisenberg picture that generate a linear relationship between of the annihilation operators of each mode.  That is, only those unitary operators $\mathcal{U}$ which act on the Fock basis such that
\begin{equation}
	\mathcal{U}^\dagger a_i \mathcal{U} = \sum_{j} u_{ij} a_j
\end{equation}
where $a_i$ is the $i$-th mode's annihilation operator and the $u_{ij}$ form a unitary matrix which for $m$ modes is a $m \times m$ matrix.  It is important to make a distinction from the unitary operator $\mathcal{U}$ which acts upon the Fock basis and the unitary matrix defined by $u_{ij}$ which describes the linear mixing of modes.  For optical systems the matrix $u_{ij}$ is determined by how linear optical elements, such as beam-splitters and phase shifters, are laid out.  In fact all unitary networks can be constructed using just beam-splitter and phase shifters~\cite{Reck94}.

\begin{figure*}
	\includegraphics[width=129mm]{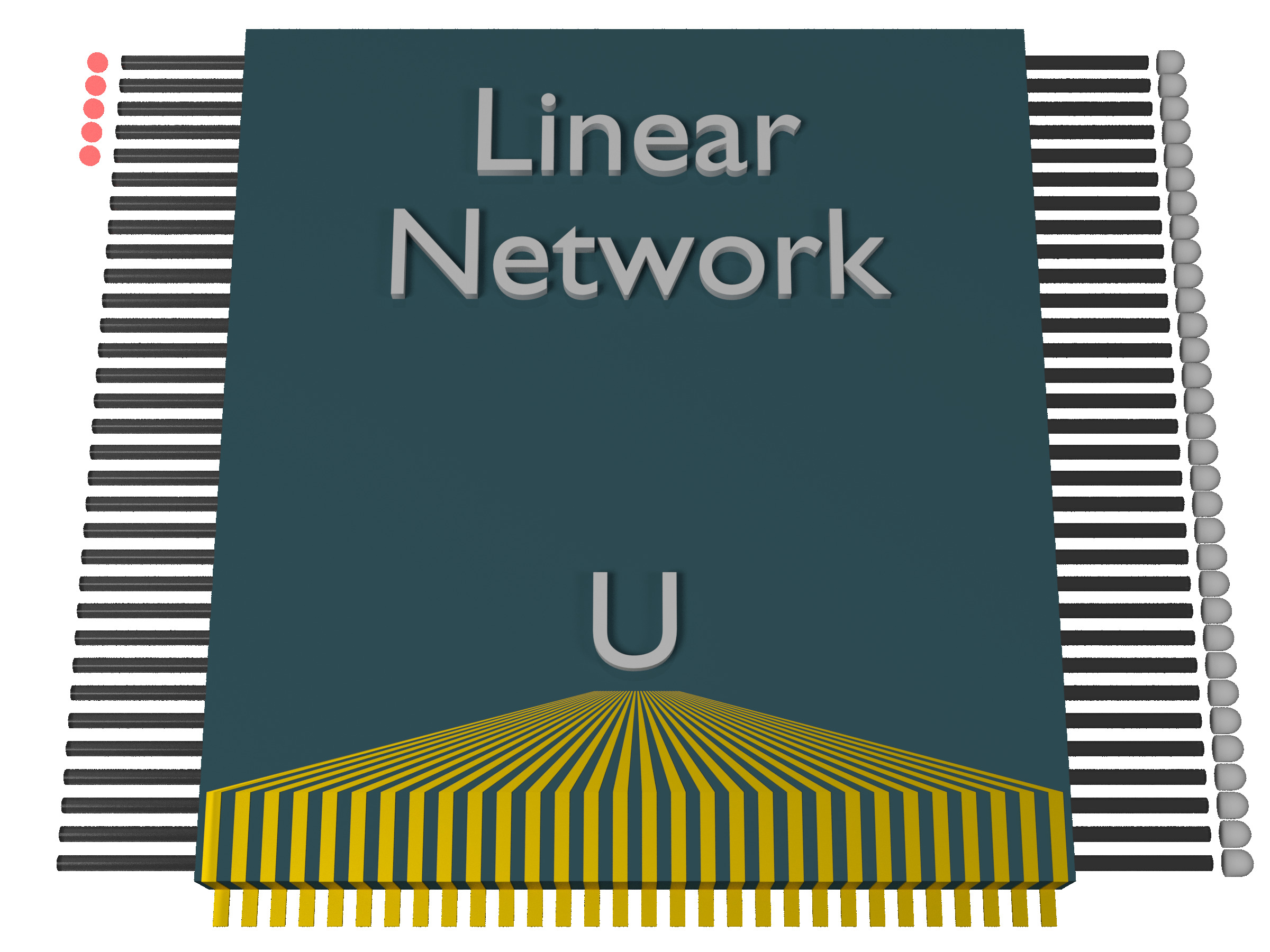}
	\caption{\label{bsamp}Schematic representation of a 5 photon, 32 mode instance of the $\BosonSampling$ problem.  The photons are injected individually into the input modes (left), interacted linearly through a linear network that has scattering matrix $u$ which is classically controlled (bottom) and all outputs are detected in the Fock basis (right). }
\end{figure*}

The class $\BosonSampling$ is defined as quantum sampling problems where a fiducial $m$-mode $n$-boson Fock state
\begin{equation}
	\label{eq:bsinput}
	\ket{\underbrace{1,1,1,\ldots,1}_n,\underbrace{0,0,\ldots,0}_{m-n}}
\end{equation}
is evolved through a linear network with the output being samples from the distribution that results after a Fock basis measurement of all modes.   The linear interaction is then the input to the algorithm and the output is the sample from the probability distribution.  FIG.~\ref{bsamp} shows a schematic representation of this configuration.  The set of events which are then output by the algorithm is a tuple of $m$ non-negative integers whose sum is $n$.  This set is denoted $\Phi_{m,n}$.

The probability distribution of output events is related to the matrix permanent of sub-matrices of $u_{ij}$.  The matrix permanent is defined in a recursive way like the common matrix determinant, but without the alternation of addition and subtraction.  For example
\begin{equation}
	Per \left(\begin{array}{cc}a & b \\ c & d\end{array}\right) = a d + c b.
\end{equation}
\begin{equation}
	Per \left(\begin{array}{ccc}a & b & c \\ d &e &f \\g &h &i \end{array}\right) = a e i + a h f + b d i + d g f + c d h + c g e .
\end{equation}
Or in a more general form
\begin{equation}
	Per (A) = \sum_{\sigma \in \mathcal{S}_n} \prod_{i=1}^n a_{i,\sigma(i)}
\end{equation}
where $\mathcal{S}_n$ represents the elements of the symmetric group of permutations of $n$ elements.  With this, we can now define the output distribution of the linear network with the input state from Eq.~\ref{eq:bsinput}.  For an output event $S=(s_1,s_2,\ldots,s_n)\in \Phi_{m,n}$, the probability of $S$ is then
\begin{equation}
	\label{eq:BSprobability}
	p_S = \frac{|Per(A_S)|^2}{s_1! s_2! \ldots s_n!}
\end{equation}
where the matrix $A_S$ is a $n \times n$ sub-matrix of $u_{ij}$ where row $i$ is repeated $s_i$ times and only the first $n$ columns are used.  One critical observation of this distribution is that all events are proportional to the square of a matrix permanent derived from the original network matrix $u_{ij}$.  Also, the fact that each probability is derived from a permanent of a sub-matrix of the same unitary matrix ensures all probabilities are less than $1$ and the distribution is normalised. 

The complexity of computing the matrix permanent is known to be $\SharpP$-complete for the case of matrices with entries that are $0$ or $1$~\cite{Val79}.  It is also possible to show that for a matrix with real number entries is $\SharpP$-hard to multiplicatively estimate~\cite{AA}.  Therefore, using the argument presented above, the case of sampling from this exact probability distribution implies a polynomial hierarchy collapse.

The question is then if sampling from approximations of $\BosonSampling$ distributions also implies the same polynomial hierarchy collapse.  The answer that Aarsonson and Arkipov found~\cite{AA} is that the argument does hold because of a feature that is particular to the linear optical scattering probabilities.  When performing the estimation of the matrix permanent for exact sampling, the matrix is scaled and embedded in $u_{ij}$. The probability of one particular output event, with $n$ ones in the locations of where the matrix was embedded, is then proportional to the matrix permanent squared.  The matrix permanent can then be estimated multiplicatively in the third level of the polynomial hierarchy.  But any event containing $n$ ones in $\Phi_{m,n}$ could have been used to determine the location of the embedding.  This means that, if the estimation is made on a randomly chosen output event, and that event is hidden from the algorithm implementing approximate $\BosonSampling$, then the expected average error in the estimation will be the overall permitted error divided by the total number of events which could have been used to perform the estimation. 

An important consideration of the approximate sampling argument is that the input matrix appears to be drawn from Gaussianly distributed random matrices.  This ensures that there is a way of randomly embedding the matrix into $u_{ij}$ so that there is no information accessible to the algorithm about where that embedding has occurred.  This is possible when the unitary network matrix is sufficiently large (strictly $m=O(n^5\ln^2 n)$ but $m=O(n^2)$ is likely to be OK).  Also, under this condition, the probability of events detected with two or more bosons in a single detector tends to zero for large $n$ (the so-called ``Bosonic Birthday Paradox'').  There are $\binom{m}{n}$ events in $\Phi_{m,n}$ with only $n$ ones and so the error budget can be evenly distributed over just these events.  There are exponentially many of these events and so the error in the probability of an individual event does not dominate but is as small as the average expected probability itself.

With this assumption about the distribution of input matrices, the proof for hardness of approximate sampling relies on the problem of estimating the permanents of Gaussian random matrices still being in $\SharpP$-hard.  Furthermore, as the error allowed to the sampling probabilities is defined in terms of total variation distance, the error in estimation becomes additive rather than multiplicative.

This changes the situation from the hardness proof for exact sampling enough to be concerned that the proof may not apply.  Aaronson and Arkipov therefore isolated the requirements for the hardness proof to still apply down to two conjectures that must hold for additive estimation of permanents for Gaussian random matrices to be $\SharpP$-hard.  They are the Permenants of Gaussian Conjecture (PGC) and the Permenant Anti-Concentration Conjecture (PACC).  The PACC conjecture says that if the matrix permanents of Gaussian random matrices are not too concentrated around zero. If this holds then additive estimation of permanents for Gaussian random matrices is polynomial-time equivalent to multiplicative estimation.  The PGC is that multiplicative estimation of permanents from Gaussian random matrices is $\SharpP$-hard.  In both of these conjectures there are related proofs that seem close, but do not exactly match the conditions required. Nevertheless, both of these conjectures are highly plausible.

\section{Experimental implementations of $\BosonSampling$}

Several small scale implementations of $\BosonSampling$ have been performed with quantum optics. Implementing $\BosonSampling$ using optics is an ideal choice as the linear network consists of a large multi-path interferometer.  Then the inputs are single photon states which are injected into the interferometer and single photon counters are placed at all $m$ output modes and the arrangement of photons at the output, shot-by-shot, is recorded.  Due to the suppression of multiple photon counts under the conditions for approximate $\BosonSampling$, single photon counters can be replaced by detectors that detect the presence or absence of photons (e.g. avalanche photo diodes).

Within these optical implementations, the issues of major concern are photon loss, mode-mismatch, network errors and single photon state preparation and detection imperfections.  Some of these issues can be dealt with by adjusting the theory and checking that the hardness proof still holds.  In the presence of loss one can post-select on events where all $n$ photons make it to the outputs.  This provides a mechanism to construct proof of principle devices but does incur an exponential overhead which prevents scaling to large devices. Rohde and Ralph studied bounds on loss in BosonSampling by finding when efficient classical simulation of lossy BosonSampling is possible in two simulation strategies: Gaussian states and distinguishable input photons~\cite{Rohde2012}.   Aaronson and Brod~\cite{Aaronson2016} have shown that in the case where the number of photons lost is constant, then hardness can still be shown.  However, this is not a realistic model of loss as the number of photons lost will be proportional to the number of photons input.  Leverrier and Garcia-Patron have shown that it is a necessary condition for errors in the network to be tolerable provided the error in the individual elements scales as $O(n^{-2})$~\cite{Lev2014}.  Later Arkipov showed the sufficient condition is element errors scaling as  $o(n^{-2}\log^{-1}m)$~\cite{Arkhipov2015}. Rahimi-Keshari, et al. showed a necessary condition for hardness based on the presence of negativity of phase-space quasiprobability distributions~\cite{Rahimi2016}.  This give inequalities constraining the overall loss and noise of a device implementing $\BosonSampling$.

The majority of the initial experiments were carried out with fixed, on-chip interferometers \cite{TIL13,SPR13,CRE13,BAR13}, though one employed a partially tunable arrangement using fibre optics \cite{BRO13}. The largest network so far was demonstrated by N.~Spagnolo, et al, where 3 photons were injected into 5, 7, 9 and 13 mode optical networks~\cite{SPA14}. In this experiment the optical networks were multi-mode integrated interferometers fabricated in glass chips by femtosecond laser writing. The photon source was parametric down-conversion with four photon events identified via post-selection, where 3 of the photons were directed through the on-chip network and the 4th acted as a trigger. Single photon detectors were placed at all outputs, enabling the probability distribution to be sampled. 

For the 13 mode experiment there are 286 possible output events from $\Phi_{13,3}$ consisting of just zeros and ones.  To obtain the expected probability distributions the permanents for the sub-matrices corresponding to all configurations were calculated. Comparing the experimentally obtained probabilities with the predictions showed excellent agreement for all the chips. Such a direct comparison would become intractable for larger systems -- both because of the exponentially rising complexity of calculating the probabilities, and because of the exponentially rising amount of data needed to experimentally characterise the distribution. N.~Spagnolo, et al demonstrated an alternative approach whereby partial validation of the device can be obtained efficiently by ruling out the possibility that the distribution was simply a uniform one \cite{AAR13}, or that the distribution was generated by sending distinguishable particles through the device \cite{CAR13}. In both cases, only small sub-sets of the data were needed and the tests could be calculated efficiently.

The BosonSampling problem is interesting because, as we have seen, there are very strong arguments to suggest that medium scale systems, such as 50 bosons in 2,500 paths, are intractable for classical computers. Indeed, even for smaller systems, say 20 bosons in 400 paths, no feasible classical algorithms are currently known which can perform this simulation. This suggests that quantum computations can be carried out in this space without fault tolerant error correction that may rival the best current performance on classical computers. In addition, there is a variation of the problem referred to as scatter-shot, or Gaussian BosonSampling which can be solved efficiently by  directly using the squeezed states deterministically produced by down converters as the input (rather than single photon states) \cite{LUN14} which has been experimentally demonstrated on a small scale using up to six independent sources for the Gaussian states~\cite{Bentivegna15}. Thus the major challenge to realising an intermediate optical quantum computer of this kind is the ability to efficiently (i.e. with very low loss and noise) implement a reconfigurable, universal linear optical network over hundreds to thousands of modes. On-chip designs such as the 6 mode reconfigurable, universal circuit demonstrated by J.~Carolan, et al~\cite{CAR15} are one of several promising ways forward. Another interesting approach is the reconfigurable time-multiplexed interferometer proposed by Motes et al.~\cite{Motes2014} and recently implemented in free-space by Y.~He, et al~\cite{HE16}. This latter experiment is also distinguished by the use of a quantum dot as the single photon source which have also be utilised in the spatial multiplexed interferometers~\cite{Loredo16}.  Finally, it is possible to construct a theory for realistic interferometers including polarisation and temporal degrees of freedom can be considered and also give rise to probabilities proportional to matrix permanents~\cite{Laibacher15, Tamma16}.

\section{Sampling with the circuit model and $\IQP$}

Last year Bremner, Montanaro, and Shepherd extended the $\BosonSampling$ argument to $\IQP$ (Instantaneous Quantum Polynomial-time) circuits, arguing that if such circuits could be classically simulated to within a reasonable additive error, then the Polynomial Hierarchy would collapse to the third level \cite{BMS15}. Crucially, these hardness results rely only on the conjecture that the average-case and worst-case complexities of quantum amplitudes of $\IQP$ circuits coincide.  Only the one conjecture is needed  as the $\IQP$ analogue of the PACC was proven to be true. As this argument is native to the quantum computing circuit model, any architecture for quantum computation can implement $\IQP$ Sampling. It also means that error correction techniques can be used to correct noise in such implementations. Furthermore, the $\IQP$ Sampling and the related results on Fourier Sampling by Fefferman and Umans \cite{Fefferman15} demonstrate that generalizations of the Aaronson and Arkhipov argument \cite{AA} could potentially be applied to a much wider variety of quantum circuit families, allowing the possibility of sampling arguments that are both better tailored to a particular experimental setup and for their complexity to be dependent on new theoretical conjectures.

\begin{figure}
	\includegraphics[width=86mm]{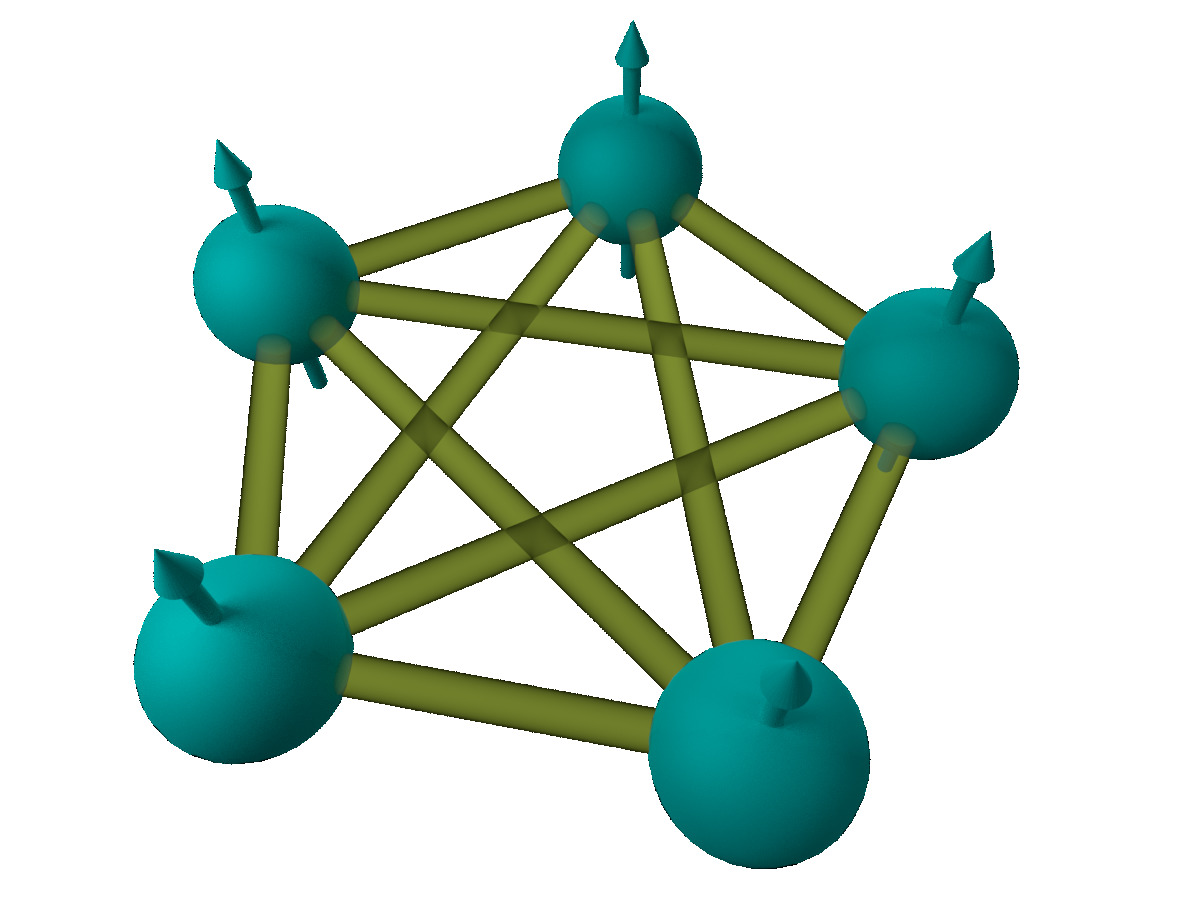}
	\caption{\label{IQPfig}Five qubit random Ising model with commuting $X \otimes X$ interactions with random strengths is an example of a problem within the class $\IQP$.  Qubits are prepared and measured in the computational basis.}
\end{figure}

$\IQP$ circuits \cite{BS09,BJS10} are an intermediate model of quantum computation where every circuit has the form $C=H^{\otimes n}DH^{\otimes n}$, where $H$ is a hadamard gate and $D$ is an efficiently generated quantum circuit that is diagonal in the computational basis. \emph{\IQP sampling} then simply corresponds to performing measurements in the computational basis on the state $H^{\otimes n}DH^{\otimes n}|0\rangle^{\otimes n}$. In \cite{BMS15} it was argued that classical computers could not efficiently sample from $\IQP$ circuits where $D$ is chosen uniformly at random from circuits composed of: (1) $\sqrt{CZ}$ (square-root of controlled-Z), and $T=\left( \begin{smallmatrix} 1&0\\0 & e^{i\pi/4} \end{smallmatrix} \right)$ gates; or (2) $Z$, $CZ$, and $CCZ$ (doubly controlled-Z) gates. This argument was made assuming that it is $\SharpP$-hard to multiplicatively approximate a constant fraction of instances of (the modulus-squared of): (C1) the complex-temperature partition function of a random 2-local Ising model; or (C2) the (normalized) gap of a degree-3 polynomial (over $\mathbb{F}_2$). These conjectures can be seen as $\IQP$ analogues of Boson Sampling's PGC. In the case of (1) these circuits correspond to random instances of the Ising model drawn from the complete graph,  as depicted in FIG.~\ref{IQPfig}. 

The worst-case complexity of the problems in both (C1) and (C2) can be seen to be $\SharpP$-hard as these problems are directly proportional to the output probabilities of the $\IQP$ circuit families (1) and (2). These families are examples of sets that become universal under postselection and as a result their output probabilities are $\SharpP$-hard (as mentioned in Section \ref{SecSampling}). This is shown by noting that for either of the gate sets (1) or (2), the only missing ingredient for universality is the ability to perform hadamard gates at any point within the circuit. In \cite{BJS10} it was shown that such gates can be replaced with a ``hadamard gadget", which requires 1 postselected qubit and controlled-phase gate per hadamard gate. It can be shown that the complexity of computing the output probabilities of IQP circuits, $p_x= |\langle x|H^{\otimes n}DH^{\otimes n}|0\rangle^{\otimes n}|^2$, is $\SharpP$-hard in the worst case and this also holds under multiplicative approximation \cite{BMS15, Fujii13, Goldberg14}. 

The hardness of $\IQP$-sampling to within additive errors follows from the observation that Stockmeyer's algorithm combined with sufficiently accurate classical additive simulation returns a very precise estimate to the probability $p_0=|\langle 0|^{\otimes n}C_y|0\rangle^{\otimes n}|^2$ for a wide range of randomly chosen circuits $C_y$. A multiplicative approximation to $p_0$ can be delivered on a large fraction of choices of $y$ when both: (a) for a random bitstring $x$, the circuit $\bigotimes_{i=1}^nX^{x_i}$ is a hidden subset of the randomly chosen circuits $C_y$; and (b) $p_0$ anti-concentrates on the random choices of circuits $C_y$. Both of these properties hold for the randomly chosen $\IQP$ circuit families (1) and (2) above, and more generally hold for any random family of circuits that satisfies the Porter-Thomas distribution \cite{Boixo16}. Classical simulations of samples from $C_y$ implies a Polynomial Hierarchy collapse if a large enough fraction of $p_0$ are also $\SharpP$-hard under multiplicative approximations - and definitively proving such a statement remains a significant mathematical challenge. As mentioned above in \cite{BMS15} the authors could only demonstrate  sufficient \emph{worst-case} complexity for evaluating $p_0$ for the circuit families (1) and (2), connecting the complexity of these problems to key problems in complexity theory. 

The $\IQP$ circuit families discussed above allow for gates to be applied between any qubits in a system. This means that there could be $O(n^2)$ gates in a random circuit for (1) and $O(n^3)$ gates for (2), with many of them long-range. From an experimental perspective this is challenging to implement as most architectures have nearest-neighbour interactions. Clearly these circuits can be implemented with nearest-neighbour gates from a universal gate set, however many SWAP gates would need to be applied. Given that many families of quantum circuits can have $\SharpP$-hard output probabilities this suggests it is worthwhile understanding if more efficient schemes can be found. It is also important to identify new average-case complexity conjectures that might lead to a proof that quantum computers cannot be classically simulated. 

The challenge in reducing the resource requirements for sampling arguments is to both maintain the anti-concentration property and the conjectured $\SharpP$-hardness of the average-case complexity of the output probabilities. Recently it was shown that \emph{sparse} $\IQP$-sampling, where $\IQP$ circuits are associated with random sparse graphs, has both of these features \cite{BMS16}. It was proved that anticoncentration can be achieved with only $O(n\log n)$ long-range gates or rather in depth $O(\sqrt{n} \log n)$ with high probability in a universal 2d lattice architecture. 

If we take as a guiding principle that in the worst-case output probabilities should not have a straightforward sub-exponential algorithm, then the 2d architecture depth cannot be less than $O(\sqrt{n})$ as there exist classical algorithms for computing any quantum circuit amplitude for a depth $t$ circuit on a 2d-lattice that scale as $O(2^{t\sqrt{n}})$. This suggests that there might be some room still to optimize the results of \cite{BMS16}, and is further evidenced by a recent numerical study suggesting that anti-concentration, and subsequently quantum supremacy, could be achieved in systems where gates are chosen at random from a universal gate set on a square lattice with depth scaling like $O(\sqrt{n})$. Such arguments give hope that a quantum experiment on approximately 50 qubits could be performed, assuming that the rate of error can be kept low enough.

Recently it has also been proposed that sampling from 2d ``brickwork'' states cannot be classically simulated \cite{Gao16}. Such states have depth $O(1)$ and as such their output probabilities are thought not to anticoncentrate and can be classically computed in sub-exponential $2^{O(\sqrt{n})}$ time.  However, the authors argue that there are some output probabilities that are $\GapP$-complete, yet might be reliably approximated via Stockmeyer's algorithm without anticoncentration. This is possible under considerably stronger average-case complexity conjectures than those appearing in \cite{BMS15, BMS16}, and also requires polynomially more qubits.

Finally it should be remarked that the level of experimental precision required to definitively demonstrate quantum supremacy, even given generous constant total variation distance bound (such as required in \cite{BMS15}), is very high. Asymptotically this typically requires the precision of each circuit component must improve by an inverse polynomial in the number of qubits. This is likely hard to achieve with growing system size without the use of fault tolerance constructions. More physically reasonable is to assume that each qubit will at least have a constant error rate, which corresponds to a total variation distance scaling like $O(n)$. Recently it was shown that if an $\IQP$ circuit has the anti-concentration property, and it suffers from a constant amount of depolarizing noise on each qubit then there is an classical algorithm that can  classically simulated it to within a reasonable total variation distance \cite{BMS16}. However, it should be remarked for a constant number of qubits this algorithm will likely still have a very large run-time. By contrast, $\IQP$ remains classically hard under the error model for multiplicative classical approximations \cite{Fujii16}. Intriguingly, this class of errors can be corrected without the full arsenal of fault tolerance, retrieving supremacy for additive error approximations requiring only operations from $\IQP$ albeit with a cost in terms of gates and qubits \cite{BMS16}. This suggests that unambiguous quantum supremacy may yet require error correction, though the level of error correction required remains a very open question.

\section{Conclusion}

Quantum sampling problems have provided a path towards experimental demonstration of the supremacy of quantum algorithms with significantly lower barriers than previously thought necessary for such a demonstration.   The two main classes of sampling problems demonstrating quantum supremacy are $\BosonSampling$ and $\IQP$ which are intermediate models of optical and qubit based quantum information processing architectures.  Even reasonable approximations to the outputs from these problems, given some highly plausible conjectures, are hard for classical computers to compute.  

Some future directions for research in this area involve a deeper understanding of these classes as well as experimentally addressing the technological challenges towards implementations that outperform the current best known classical algorithms.  Theoretical work on addressing what is possible within these classes, such as detecting and correcting with errors within the intermediate models will both aid understanding and benefit experimental implementations.  There has been some study of the verification of limited aspects of these devices~\cite{Tichy14,Walschaers16,Aolita15} but more work is required.  As $\BosonSampling$ and $\IQP$ are likely outside the Polynomial Hierarchy, an efficient reconstruction of the entire probability distribution which is output from these devices will likely be impossible.  However, one can build the components, characterise them and their interactions, build and run such a device to within a known error rate. Beyond this multiplayer games based on sampling problems in $\IQP$ have been proposed to test whether a player is actually running an $\IQP$ computation \cite{BS09}.  Recently the complexity of $\IQP$ sampling has been connected to the complexity of quantum algorithms for approximate optimization problems \cite{Farhi16}, suggesting further applications of $\IQP$ and closely related classes. Applications of BosonSampling to molecular simulations~\cite{Huh2015}, metrology~\cite{Motes2015} and decision problems~\cite{Nikolopoulos2016} have been suggested, though more work is needed in this space.  Nevertheless, the results from quantum sampling problems have undoubtedly brought us closer to the construction of a quantum device which definitively displays the computational power of quantum mechanics.

\section*{Contributions}

All authors contributed equally to this work.

\section*{Funding}

APL and TCR received financial support from the Australian Research Council Centre of Excellence for Quantum Computation and Communications Technology (Project No. CE110001027). MJB has received financial support from the Australian Research Council via the Future Fellowship scheme (Project No. FT110101044). 

\section*{Competing interests}

The authors declare no competing financial interests.

\end{document}